\documentclass[a4paper,11pt,nohyper]{JHEP3}
\usepackage{amsmath,latexsym,amssymb,slashed}
\usepackage{graphicx}
\usepackage{psfrag}
\usepackage[latin1]{inputenc}
\usepackage{subfigure}

\newcommand\beal{\begin{align}}

\newcommand{\eq}[1]{\begin{equation}#1\end{equation}}
\newcommand{\spl}[1]{\begin{split}#1\end{split}}

\newcommand{\arXividhepth}[1]{\href{http://arxiv.org/abs/#1}arXiv:{\tt #1} [hep-th]}

\newcommand{\mcal}{\mathcal{M}}

\newcommand{\ncal}{\mathcal{N}}

\def\d{\text{d}}

\def\slashchar#1{\setbox0=\hbox{$#1$}           
\dimen0=\wd0                                 
\setbox1=\hbox{/} \dimen1=\wd1               
\ifdim\dimen0>\dimen1                        
\rlap{\hbox to \dimen0{\hfil/\hfil}}      
#1                                        
\else                                        
\rlap{\hbox to \dimen1{\hfil$#1$\hfil}}   
/                                         
\fi}

\title{Supersymmetric AdS vacua and separation of scales}

\author{Dimitrios Tsimpis\\
Universit\'{e} de Lyon\\
UMR 5822, CNRS/IN2P3, Institut de Physique Nucl\'{e}aire de Lyon\\ 
4 rue Enrico Fermi,  
F-69622 Villeurbanne Cedex, France\\
\email{tsimpis@ipnl.in2p3.fr}}

\abstract{The moduli space of the supersymmetric massive IIA AdS$_4\times S^2(\mathcal{B}_4)$ vacua, where $S^2(\mathcal{B}_4)$ is a two-sphere bundle over a four-dimensional K\"{a}hler-Einstein base $\mathcal{B}_4$, includes three independent parameters which can be thought of as corresponding to the 
sizes of AdS$_4$, $\mathcal{B}_4$ and the $S^2$ fiber. It might therefore be expected that these vacua do not suffer from the absence of scale separation. 
We show that the independence of the geometric moduli survives flux quantization. However, we uncover an attractor behavior whereby all sizes flow to equality in some neighborhood of spacetime independently of the initial conditions set by the parameters of the solution. This is further confirmed by the study of the ratio of internal to external scalar curvatures. We also show that the asymptotic Kaluza-Klein spectrum of a ten-dimensional massive scalar is governed by a scale of the order of the AdS$_4$ radius. Furthermore we point out that the curvature ratio in supersymmetric IIA AdS$_4$ vacua with rigid SU(3) structure is of order one, indicating the absence of scale separation in this large class of vacua.}

\begin{document}
\setcounter{footnote}{0}
\renewcommand{\thefootnote}{\arabic{footnote}}
\setcounter{section}{0}
\vfill\break

\section{Introduction}

\label{introduction}

There are by now several known supersymmetric AdS$_4$ string theory vacua which are pure-flux (i.e. do not contain extra ingredients such as orientifolds). These vacua are in principle very well controlled to the extent that they can be defined nonperturbatively as quantum gravity theories via a dual three-dimensional CFT. 
On the other hand, supersymmetric backgrounds of the form AdS$\times\mcal$ typically have the property that the radius of curvature $L$ of the  AdS  space is of the same order as the typical size $L_{\mathrm{int}}$ of the internal manifold $\mcal$. In the following I will refer to this feature as the absence of a separation of scales.

The absence of a separation of scales 
makes it difficult to use supersymmetric pure-flux AdS$_4$ vacua as a starting point for realistic compactifications. Schematically, for an internal space of typical size $L_{\mathrm{int}}$ there is an indexing by integers $n$ such that the masses $m_n$ of the Kaluza-Klein (KK) modes behave 
 as\footnote{The relation between the eigenvalues of 
the Laplacian of the internal space (which correspond to the squares of the masses of the KK modes) and the typical size of the internal space is mathematically more subtle than is suggested in eq.~(\ref{i1}). A precise estimate of asymptotic scaling of eigenvalues is given in sections \ref{kk} and \ref{sl}.}
\eq{\label{i1}
m_n^2\sim\frac{n^2}{L_{\mathrm{int}}^2}\sim \frac{n^2}{L^2}
~,}
where the last approximation is a consequence of the absence of scale separation. 
Hence there is no low-energy limit in which all but a finite subset of the KK modes decouple: in order to have $m^2_n\rightarrow\infty$ the AdS$_4$ space must collapse to zero size.

A related observation is the following: Uplifting the AdS$_4$ vacuum to the observed positive cosmological constant requires a quantum effect of the order of the curvature of AdS$_4$, which in the absence of a separation of scales is of the same order as the KK scale. Thus a large quantum correction is required, which means that the AdS$_4$ vacuum cannot be used as a controlled approximation of the true vacuum.

In \cite{lt3} we constructed massive  IIA $\ncal=2$ supersymmetric supergravity solutions of the form AdS$_4\times S^2(\mathcal{B}_4)$, where $S^2(\mathcal{B}_4)$ is a two-sphere bundle over a four-dimensional K\"{a}hler-Einstein base $\mathcal{B}_4$. The moduli space of these solutions includes three independent parameters which can roughly be thought of as corresponding to the sizes of AdS$_4$, $\mathcal{B}_4$ and the $S^2$ fiber. One  might therefore  expect that these supersymmetric vacua do not suffer from the absence of  scale separation, since the sizes of the  AdS$_4$ and the internal space appear to be independent. The main motivation of the present paper was to examine whether or not this is the case. 
As we will see, our results do not allow us to make a clear-cut case for scale separation, or absence thereof, in this class of vacua.

The plan of the remainder of the paper is as follows. 
In section \ref{sec2} we start by reviewing the solutions of \cite{lt3}. 
As already mentioned, these solutions are parameterized by independent parameters controlling the initial sizes of the internal and external spaces. However, we uncover an attractor behavior whereby all sizes of the ten-dimensional geometry flow to equality in some neighborhood of spacetime independently of the initial conditions set by the parameters of the solution. 

In section \ref{sec:curv} we study the ratio of internal to external scalar curvatures and we find it to be consistent with the aforementioned attractor behavior: the ratio flows to a value of order one in some neighborhood of spacetime independently of the initial parameters. As a side remark, we also show that the ratio of curvatures is of  order one in all  supersymmetric AdS$_4$ vacua of the type of \cite{lt}. This constitutes strong evidence for the absence of scale separation in this large class of supersymmetric vacua.

In section \ref{sec:flux} we show that the independence of the parameters controlling the initial sizes of the external and internal spaces survives flux quantization. In section \ref{kk} we study the KK spectrum of a ten-dimensional massive scalar in the geometry of the solutions of \cite{lt3}. The internal part of the wave equation is reduced to a singular Sturm-Liouville problem and the asymptotic KK spectrum is shown to be governed by a scale of the order of the AdS$_4$ radius. Our conclusions are summarized in section \ref{conclusions}. 
Further technical details can be found in the appendices.

\section{Massive AdS$_4\times S^2(\mathcal{B}_4)$ vacua}\label{sec2}

The solutions of \cite{lt3} can be thought of as massive IIA deformations 
of the $\ncal=2$ IIA circle reductions of 
the M-theory  AdS$_4\times Y^{p,q}(\mathcal{B}_4)$ backgrounds of \cite{gm,ms2}, where $Y^{p,q}(\mathcal{B}_4)$ is a seven-dimensional Sasaki-Einstein manifold. 
The first such massive deformation was constructed in \cite{pz} for the $\ncal=2$ IIA circle reduction of the M-theory AdS$_4\times Y^{3,2}(\mathbb{CP}^2)$ background (the $Y^{3,2}(\mathbb{CP}^2)$ 
space is also referred to as $M^{1,1,1}$ in the physics literature).

Let us begin by summarizing some relevant facts about the solutions of \cite{lt3}; some additional details can be found in section \ref{details}. These are  $\ncal=2$ warped AdS$_4\times \mcal_6$ type IIA supergravity solutions whose metric (in the string frame) 
is given by\footnote{\label{f2}
For the purposes of the 
present paper we are using different conventions than in \cite{lt3}. More specifically, the warp functions $A$, $C$ here are related to those of \cite{lt3} through:
\eq{
e^{2A_{\mathrm{here}}}= e^{2A_{\mathrm{there}}}~,~~
~e^{2C_{\mathrm{here}}}=\frac{1}{L}~\!e^{2C_{\mathrm{there}}}~.
}
The metric in eq~.(\ref{3}) is obtained from eq.~(2.17) of \cite{lt3} as follows: By fixing the reparameterization invariance we may choose the coordinate $t$ so that  
\eq{\label{a2}
e^{A-B}=W~,~~~\frac{\d}{\d t}\log\tan\varepsilon(t)=-2\tan\theta(t)~,
}
where the parameter $W$ in that reference is the inverse of the radius $L$ of AdS$_4$. Then, the penultimate line of eq.~(2.19) of \cite{lt3} gives the relation between the differentials $\d t$ and $\d\theta$: 
\eq{
f(\theta)~\!\d\theta=\d t
~,}
with $f(\theta)$ as in (\ref{4}). 
Substituting the equation above in eq.~(2.17) of \cite{lt3}, one obtains precisely eq.~(\ref{3}) of the present paper.
}
\eq{\label{1}
\d s^2_{10}=
e^{2A(\theta)}\d s^2(AdS_4)+L^{2}\d s^2(\mcal_6)
~.}
The metric $\d s^2(AdS_4)$ is that of a four-dimensional 
anti de Sitter space of radius $L$,\footnote{There is an ambiguity in the definition of the AdS$_4$ radius: since the latter enters the ten-dimensional metric through the combination $Le^{A}$, a rescaling of the  AdS$_4$ radius can be offset by an inverse rescaling of $e^A$. This ambiguity can be eliminated as in the present paper: by demanding that the  AdS$_4$ radius $L$ does not appear in the coupled system of differential equations (\ref{7}).} 
%
%
so that the scalar curvature is related to the radius through: $R=-12L^{-2}$. 
The metric of the internal space is given by
\eq{\label{3}
\d s^2(\mcal_6)=e^{2C(\theta)}\d s^2(\mathcal{B}_4)+e^{2A(\theta)}\big(
f^2(\theta)\d\theta^2+\sin^2\!\theta~\!(\d\psi+\mathcal{A})^2\big)
~,}
where
\eq{\label{4}
f(\theta):=\frac{1}{2-\sin^2\!\theta ~\!e^{2(A(\theta)-C(\theta))}}~.}
The coordinates $(\psi,\theta)$, with ranges 
$0\leq\psi< \pi$, $0\leq\theta\leq\pi$, parameterize a smooth $S^2$ fiber over $\mathcal{B}_4$; the coordinate $\psi$ parameterizes an $S^1$ fiber in the anticanonical bundle of $\mathcal{B}_4$. 
The metric $\d{}s^2(\mathcal{B}_4)$ of the four-dimensional K\"{a}hler-Einstein  space $\mathcal{B}_4$ is normalized so that\footnote{
The normalization in eq.~(\ref{5}) has the 
property that the six-dimensional  cone over the five-dimensional Sasaki-Einstein space 
\eq{
\d s^2_{SE}=\frac{1}{3}\d s^2(\mathcal{B}_4)+\frac{4}{9}(\d\psi+\mathcal{A})^2~,
}
is Ricci-flat. We note the difference from the normalization of eq.~(A.8) of \cite{lt3}: the metric $\d s^2(\mathcal{B}_4)$, the coordinate $\psi$ and the connection $\mathcal{A}$ in the present paper are related to those of \cite{lt3} through
\eq{
\d s^2(\mathcal{B}_4)^{\mathrm{here}}=3\d s^2(\mathcal{B}_4)^{\mathrm{there}}~,~~
~\psi^{\mathrm{here}}=\frac{3}{2}\psi^{\mathrm{there}}~,~~~
\mathcal{A}^{\mathrm{here}}=\frac{3}{2}\mathcal{A}^{\mathrm{there}}
~.}
}
\eq{\label{5}
R^{KE}_{mn}=2g^{KE}_{mn}~,}
while the $U(1)$ connection $\mathcal{A}$ on $\mathcal{B}_4$ is related to the K\"{a}hler form $J$ on $\mathcal{B}_4$ through
\eq{\label{6}
\d\mathcal{A}=-J
~.}
The dependence of the functions  $A$, $C$ on the coordinate $\theta$ is given implicitly through the following system of two coupled first-order differential equations:\footnote{The system of eqs.~(\ref{7}) is obtained from eq.~(2.19) of \cite{lt3} as follows: We fix the reparameterization invariance as in (\ref{a2}) of footnote \ref{f2}. 
Moreover, differentiating the first line of eq.~(2.19)  of \cite{lt3} we obtain an expression for $\d A(\theta(t))/\d t$ which reduces to the first line of (\ref{7}) of the present paper, upon taking into account the 
expression for $\d\theta/\d t$ (the penultimate line in eq.~(2.19) of \cite{lt3}). 
Similarly, the second line of (\ref{7}) of the present paper is obtained from the last line  in eq.~(2.19) of \cite{lt3} by eliminating the parameter $\varphi$, using eq.~(2.16) of that reference.}
\eq{\spl{\label{7}
A'&=\frac{1}{2}\tan\theta\frac{1-\sin^2\!\theta ~\!e^{2(A-C)}}{2-\sin^2\!\theta ~\!e^{2(A-C)}}\\
C'&=\frac{1}{4}\sin(2\theta)\frac{
e^{2(A-C)}}{2-\sin^2\!\theta ~\!e^{2(A-C)}}
\frac{1+e^{8A}}{1+\cos^2\theta e^{8A}}
~,}}
where a prime denotes differentiation with respect to $\theta$.

The system (\ref{7}) of two coupled first-order differential equations has not been solved analytically to date. On general grounds,  for a given set $(A_0,C_0)$ of `initial conditions'
\eq{
A_0:=A(\theta=0)~,~~~C_0:=C(\theta=0)
~,}
we expect a unique solution at least in a neighborhood of $\theta=0$. 
On the other hand, by virtue of (\ref{3}), we expect that the parameters  $(A_0,C_0)$ of the solution should control the size of $S^2$, $\mathcal{B}_4$ respectively. As we will see however in the following, this expectation is not entirely correct.

Eqs.~(\ref{7}) enjoy a form of `time-reversal' symmetry, where the angle $\theta$ plays the role of `time'. Indeed if we set 
$\tilde{A}(\tilde{\theta}):=A(\theta)$, 
$\tilde{C}(\tilde{\theta}):=C(\theta)$, with $\tilde{\theta}:=\pi-\theta$, then   $\tilde{A}(\tilde{\theta})$, $\tilde{C}(\tilde{\theta})$ obey the same 
equations (\ref{7}) as functions of $\tilde{\theta}$. As a consequence the curves $A(\theta)$, $C(\theta)$ we obtain by imposing initial conditions $A_0$, $C_0$ at $\theta=0$ and the curves $\tilde{A}(\tilde{\theta})$, $\tilde{C}(\tilde{\theta})$ we obtain by imposing the same initial conditions $A_0$, $C_0$ at $\tilde{\theta}=0$ (i.e. at $\theta=\pi$) are mirror images of each other with respect to the $\theta=\pi/2$ axis.

Another consequence of (\ref{7}) is the existence of the lower bound:
\eq{\label{bound}
f(\theta)\geq \frac{1}{2}
~,}
which will prove important in the following. In order to prove (\ref{bound})  it will suffice  
to restrict $\theta$ to the interval $[0,\pi/2]$; our argument can be 
readily extended to $[\pi/2,\pi]$. Moreover we have $f(0)=1/2$ and $f(\pi/2)=1$, as follows from (\ref{attr}) below. Hence (\ref{bound}) is 
satisfied at the endpoints and we may further 
restrict $\theta$ to the interior of the interval: $\theta\in(0,\pi/2)$. 
Next, we note that  
 if $A\leq C$ then $0\leq\sin^2\!\theta ~\!e^{2(A-C)}\leq 1$ and so (\ref{bound}) holds. 
Let us therefore consider the case $A>C$. 
From (\ref{7}) we compute:
\eq{\label{deriva}
\left[ \sin^2\!\theta ~\!e^{2(A-C)}\right]'=
\sin(2\theta) ~\!e^{2(A-C)}\left[1 
-\frac{1}{2}\tan^2\theta~\!\frac{(1+\alpha^2)~\!e^{2(A-C)}-1}{2-\sin^2\!\theta ~\!e^{2(A-C)}}\right]
~,}
where   
\eq{
\alpha^2:=\frac14\frac{\sin^2(2\theta)~\!e^{8A}}{1+\cos^2\theta~\!e^{8A}}
\geq0~.}
Furthermore note that 
$\sin^2\!\theta ~\!e^{2(A-C)}$ vanishes at $\theta=0$ and so, by continuity, in a right neighborhood of $\theta=0$ we will have $0\leq\sin^2\!\theta ~\!e^{2(A-C)}<2$ thus (\ref{bound}) holds in that neighborhood. We will now show that $\sin^2\!\theta ~\!e^{2(A-C)}$ is strictly smaller than $2$ for $\theta\in(0,\pi/2)$ and therefore (\ref{bound}) is obeyed everywhere. Indeed, suppose that $\sin^2\!\theta ~\!e^{2(A-C)}$ approaches 2 from below 
as $\theta\rightarrow\theta_0\in(0,\pi/2)$. Then the second term in the brackets on the right hand side of (\ref{deriva}) approaches positive infinity. (Recall that we are considering the case $A>C$ and therefore $e^{2(A-C)}>1$. Moreover we are restricting $\theta$ to $(0,\pi/2)$ and so $\tan\!\theta$, $\sin(2\theta)$ and $\alpha^2$ are all strictly positive). Hence the right-hand side of (\ref{deriva}) becomes negative, implying that $\sin^2\!\theta ~\!e^{2(A-C)}$ decreases as $\theta\rightarrow\theta_0$, leading to contradiction.

\vfill\break

\subsection{Attractor behavior}\label{sec3}

The system (\ref{7}) can be studied perturbatively in the 
neighborhood of $\theta=0$. The first few orders of the 
solution read\footnote{We have obtained the explicit form of the series 
 (\ref{exp1}),(\ref{exp2}) to a high order in the corresponding expansion parameters. We do not report the result here as it will not be used in the following.}
\eq{\spl{\label{exp1}
e^{2A}&=e^{2A_0}+\frac14~\!e^{2A_0}\theta^2+\mathcal{O}(\theta^4) \\
e^{2C}&=e^{2C_0}+\frac14 ~\!e^{2A_0}\theta^2+\mathcal{O}(\theta^4) 
~.}}
As expected, the perturbative solution is parameterized by the initial conditions $A_0$, $C_0$, which can be tuned freely and independently. 
Similar expansions can be derived in the neighborhood of the other endpoint $\theta=\pi$, replacing $A_0$, $C_0$ by $A(\pi)$, $C(\pi)$ respectively.

On the other hand, in the neighborhood of $\theta=\pi/2$ the perturbative solution of the system (\ref{7}) reads
\eq{\spl{\label{exp2}
e^{2A}&=e^{2a}+\mathcal{O}(\delta)\\
e^{2C}&=e^{2a}-\frac12 e^{2a}(e^{8a}+1)~\!\delta^2+\mathcal{O}(\delta^3)
~,}}
where  
\eq{
a:=A(\theta=\frac{\pi}{2})=C(\theta=\frac{\pi}{2})~,~~~\delta:=\frac{\pi}{2}-\theta~.
}
We observe the attractor behavior mentioned in the introduction, namely the fact that 
\eq{\label{attr}
\lim_{\theta\rightarrow\frac{\pi}{2}}(A-C)=0
~,}
independently of the initial conditions $A_0$, $C_0$. This has far-reaching 
consequences for the question of scale separation. For example, by inspection of the ten-dimensional metric (\ref{1}),(\ref{3}) we see that establishing a hierarchy between the scale of AdS$_4$ and that of $\mathcal{B}_4$ requires
\eq{
e^{2(A-C)}>>1
~.}
On the other hand, eq.~(\ref{attr}) guarantees that the above condition will 
be violated in a neighborhood of $\theta=\pi/2$, no matter how we tune the initial parameters $A_0$, $C_0$.

\subsection*{Numerics}

\begin{figure}
\centering
\subfigure[$e^{A_0}=1$ , $e^{C_0}=0.04$]{
\includegraphics[width=15cm,viewport=45 20 2800 900,clip]{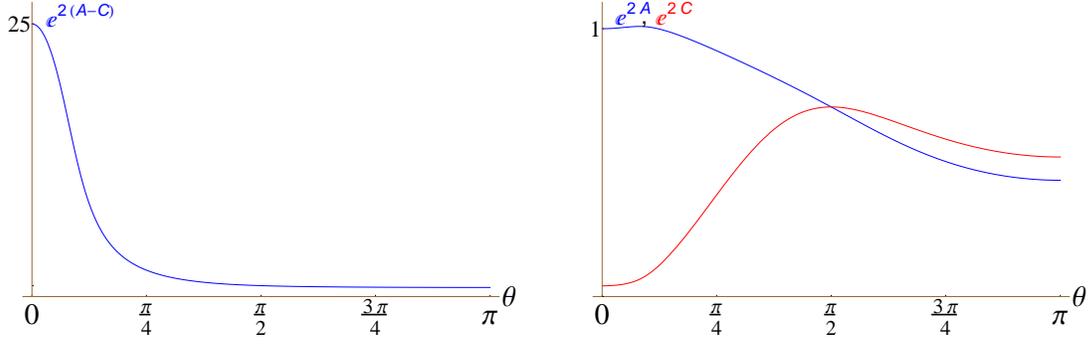}
}
\centering
\subfigure[$e^{A_0}=1$ , $e^{C_0}=10^{-2}$]{
\includegraphics[width=15cm,viewport=65 20 2670 900,clip]{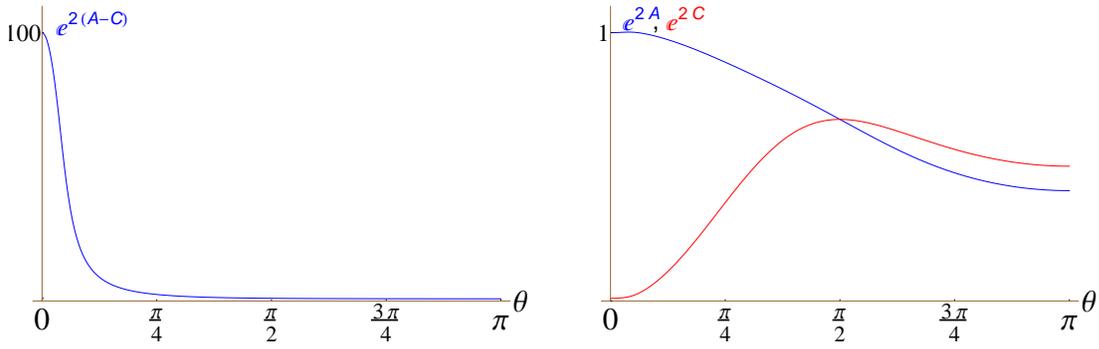}
}
\caption{The graphs of  $e^{2A}$,  $e^{2C}$,  $e^{2(A-C)}$ exhibit an attractor behavior: independently of 
the initial conditions $A_0$, $C_0$  the solution obeys  $\lim_{\theta\rightarrow\pi/2}(A-C)=0$. 
Furthermore in the above graphs we have  
$e^{2(A-C)}\lesssim\mathcal{O}(1)$ for $\pi/4\lesssim\theta$.}
\label{graph}
\end{figure}

The system (\ref{7}) can be solved numerically for given initial conditions $A_0$, $C_0$. For a large number of sample of points in the domain $10^{-4}\leq A_0, C_0\leq 10^4$, we have confirmed that 
\eq{
e^{2(A-C)}\lesssim\mathcal{O}(1)
~,~~~\mathrm{for}~~\frac{\pi}{4}\lesssim\theta\lesssim\frac{3\pi}{4}~,}
independently of the initial conditions $A_0$, $C_0$. 
Figure \ref{graph} depicts the graphs of  $e^{2A}$,  $e^{2C}$,  $e^{2(A-C)}$  for different values of $A_0$, $C_0$.
\begin{figure}
\centering
\subfigure[$e^{A_0}=1$ , $e^{C_0}=0.04$]{
\includegraphics[width=7cm,viewport=30 2 1070 600,clip]{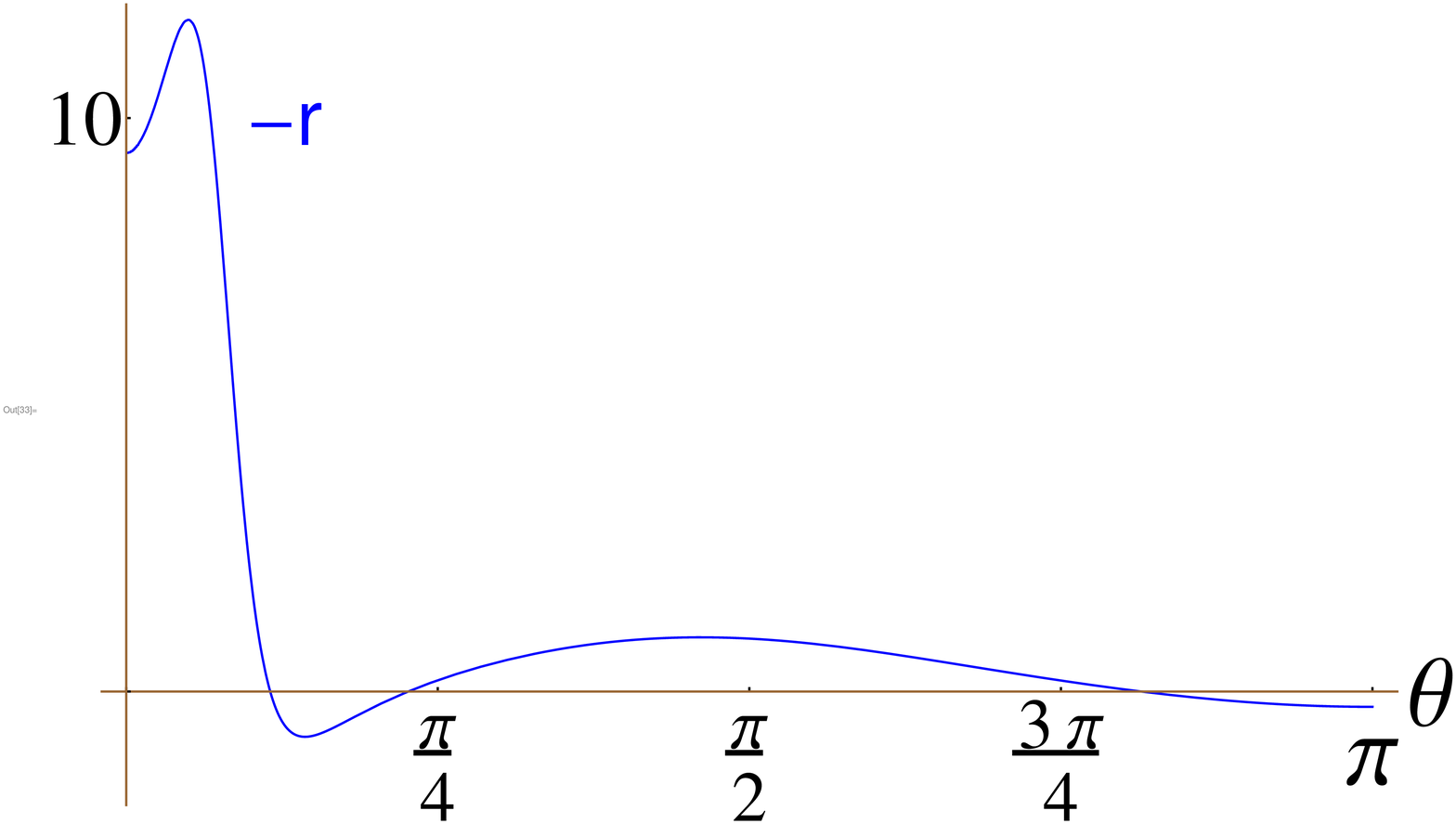}
}\hspace{0.2cm}
\subfigure[$e^{A_0}=1$ , $e^{C_0}=10^{-2}$]{
\includegraphics[width=7cm,viewport=30 2 1070 600,clip]{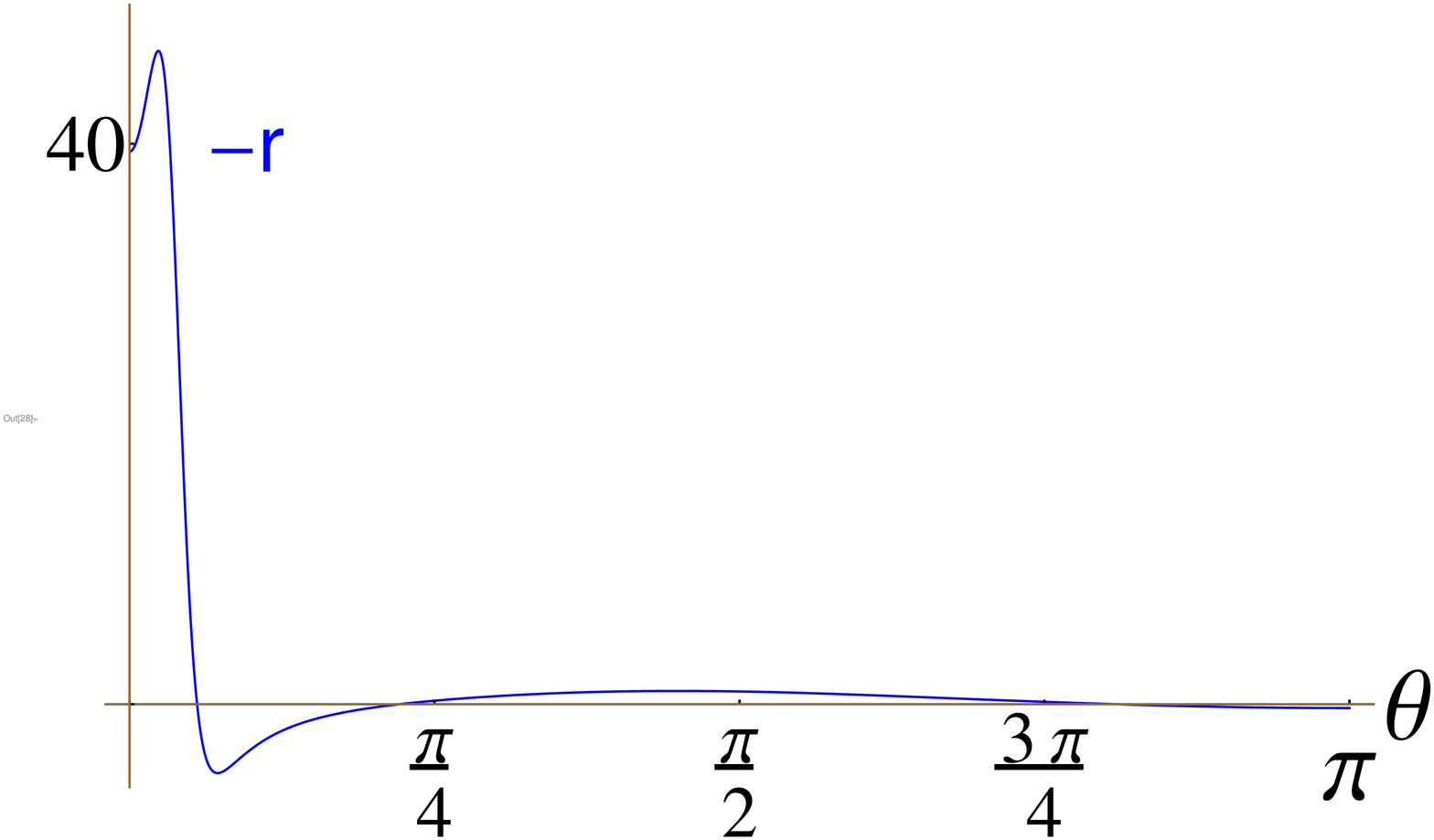}
}
\caption{The graph of the internal to external curvature ratio $r$, for 
two different sets of initial conditions $A_0$,$C_0$. The ratio obeys 
$|r|\lesssim\mathcal{O}(1)$ for $\pi/4\lesssim\theta$.}
\label{graphr}
\end{figure}

\section{Curvature ratios}\label{sec:curv}

As a consequence of the form of the  metric (\ref{1}), the ten-dimensional Ricci tensor does not contain any cross terms with one leg along the AdS$_4$ and another along the internal directions. Moreover the ratio of internal to external curvatures,
\eq{\label{ratio}
r:=\frac{R^{\mathrm{int}}}{R^{\mathrm{ext}}}
~,}
only depends on the angle $\theta$, as we will see in the following. 
The behavior of 
$r$ as a function of $\theta$ is relevant to the question 
of scale separation: If the ratio obeys $|r|>>1$ almost everywhere in the $\theta$-interval $[0,\pi]$ one could conclude that there is a separation of scales. However as we will show, although by tuning $A_0>>C_0$ we can achieve $|r|>>1$ in a neighborhood of $\theta=0$, it turns out that $|r|$ 
is always bounded in a neighborhood of $\theta=\pi/2$, independently of the 
parameters $A_0$, $C_0$.\footnote{The calculations of this section are presented for the ten-dimensional metric in the string frame. The conclusions, in particular the observed attractor behavior, are qualitatively identical for any frame.}

More specifically, for the AdS$_4$ (external) part we find:
\eq{\label{rext}
R^{\mathrm{ext}}_{\mu\nu}~\!\d x^{\mu}\otimes\d x^{\nu}=-\frac{a_0}{L^2}~\!\d s^2(AdS_4)
~,}
where
\eq{\label{a0}
a_0:=3+\frac{A''+A'(\cot\theta+4C')+4A^{\prime 2}}{f^2}-\frac{A'f'}{f^3}
~.}
Similarly, for the internal part we compute:
\eq{\label{rint}
R^{\mathrm{int}}_{mn}~\!\d x^{m}\otimes\d x^{n}=a_1~\!\d s^2(\mathcal{B}_4)
+a_2~\!(\d\psi+\mathcal{A})\otimes(\d\psi+\mathcal{A})+a_3~\!\d\theta\otimes\d\theta
~,}
where
\eq{\spl{\label{a1}
a_1&:=2-\frac{1}{2}e^{2(A-C)}\sin^2\theta-\frac{e^{2(C-A)}}{f^2}\left[
C'(\cot\theta+4A'+4C'-(\log\!f)')+C''
\right]\\
a_2&:=\frac{\sin^2\theta}{f^2}\big[
1+e^{4(A-C)}f^2\sin^2\theta\\
&~~~~~~~~~~~~~-\cot\theta~\!(5A'+4C'-(\log\!f)')
-A'~\!(4A'+4C'-(\log\!f)')-A''
\big]\\
a_3&:= 1-\cot\theta A'+4C'(A'-C')+(\log\!f)'(\cot\theta+5A'+4C')-5A''-4C''
~.}}
From the above we can compute the internal and external scalar curvatures:
\eq{\spl{\label{scalar}
L^2e^{2A}R^{\mathrm{ext}}&=-4a_0\\
L^2e^{2A}R^{\mathrm{int}}&=4e^{2(A-C)}a_1+\csc^2\!\theta ~\!a_2
+f^{-2}a_3
~.}}
These expressions are finite at $\theta=0,\pi$ (as well as in the interior 
of the $\theta$-interval $[0,\pi]$). More specifically, using (\ref{exp1}) we find:
\eq{\label{l1}
R^{\mathrm{ext}}\rightarrow -\frac{20}{L^2}~\!e^{-2A_0}~,~~~
R^{\mathrm{int}}\rightarrow \frac{1}{L^2}~\!e^{-2A_0}\big(8e^{2(A_0-C_0)}-12\big)
~,}
as $\theta\rightarrow 0$ and similarly for $\theta\rightarrow\pi$. Moreover, using expansion (\ref{exp2}) we compute:
\eq{\label{l2}
R^{\mathrm{ext}}\rightarrow -\frac{2}{L^2}~\!e^{-2a}\big(e^{8a}+5 \big)~,~~~
R^{\mathrm{int}}\rightarrow -\frac{1}{L^2}~\!e^{-2a}\big(7e^{8a}-16 \big)
~,}
as $\theta\rightarrow \pi/2$. 
It follows from the above formul\ae{} that the ratio (\ref{ratio}) of internal to external curvatures, 
only depends on the angle $\theta$. From (\ref{l1}),(\ref{l2}) we obtain the following limiting values:
\eq{\label{l3}
r\rightarrow
\left\{ \begin{array}{cl}
\frac{3}{5}\left(1-\frac{2}{3}~\!e^{2(A_0-C_0)}\right)&~,
~~\mathrm{as}~~\theta\rightarrow 0\\
\frac{7e^{8a}-16}{2e^{8a}+10}&~,~~\mathrm{as}~~\theta\rightarrow \frac{\pi}{2}   \end{array} \right.
~.}
If the ratio obeyed $|r|>>1$ almost everywhere in the $\theta$-interval $[0,\pi]$ we could conclude that there is a separation of scales. 
However, although by tuning $A_0>>C_0$ we can achieve $|r|>>1$ in a neighborhood of $\theta=0$, as can be seen from (\ref{l3}) $|r|$ is bounded 
at  $\theta=\pi/2$:
\eq{\label{rbound}
-\frac{8}{5}\leq\lim_{\theta\rightarrow\pi/2}r\leq \frac72
~.}
Moreover, we have studied the ratio $r$ numerically for a large sample of initial values $10^{-4}\leq A_0, C_0\leq 10^4$ with the conclusion that $r\lesssim\mathcal{O}(1)$ for $\pi/4\lesssim\theta\lesssim3\pi/4$. Figure \ref{graphr}, which contains the graph of $r$ for two different sets of initial conditions $A_0$, $C_0$, illustrates this behavior. The ratio of the integrated (over $\theta$) internal to external curvatures can also be seen to be order one.

\subsection{Rigid SU(3) vacua}\label{rigid}

As a side remark, 
it is interesting to note that the curvature ratio $r$ is even more 
constrained in the case of massive $\ncal=1$ IIA vacua with rigid 
SU(3) structure. The rigid SU(3) solutions of \cite{lt} are $\mathcal{N}=1$ 
supersymmetric IIA vacua of the form:
\eq{
\d s^2=\d s^2(AdS_4)+\d s^2(\mcal_6)
~,}
where $\mcal_6$ is an SU(3)-structure manifold whose only 
non-zero torsion classes are given by $W^-_1$, $W^-_2$. 
Supersymmetry relates the radius $L$ of AdS$_4$ to the torsion classes as follows:
\eq{\label{adsr}
\frac{1}{L^2}=\frac{3}{5}|W^-_1|^2-\frac{1}{80}|W^-_2|^2
~.}
Moreover, the square of the Romans mass $F_0^2$ is given by:
\eq{\label{subound}
g_s^2F_0^2=\frac{5}{16}\left(3|W^-_1|^2-|W^-_2|^2\right) 
~.}
On the other hand, the scalar curvature $R^{\mathrm{int}}$ of the SU(3)-structure manifold $\mcal_6$ can also be expressed in terms 
of the torsion classes \cite{sc}:
\eq{\label{scalarcurv}
R^{\mathrm{int}}=\frac{15}{2}|W^-_1|^2-\frac{1}{2}|W^-_2|^2
~.}
From (\ref{adsr})-(\ref{scalarcurv}), taking into account that the curvature of AdS$_4$ is $-12/L^2$, we obtain the following ratio of internal to external curvature:
\eq{
r=-\frac{4g_s^2F_0^2+|W^-_2|^2}{\frac{96}{25}~\!g_s^2F_0^2+\frac{9}{8}~\!|W^-_2|^2}
~,
}
which is bounded:
\eq{
\frac{8}{9}\leq -r\leq\frac{25}{24}
~.}
We conclude that the scale separation condition $|r|>>1$ cannot be satisfied for the 
rigid SU(3) vacua of \cite{lt}.

\section{Flux quantization}\label{sec:flux} 

We will now show that the parameters $A_0$, $C_0$ of the solution remain independent even after imposing flux quantization. 

The RR $p$-form fluxes $F_p$ of the solution are given by:\footnote{In comparison with \cite{lt3} we have reinstated an integration constant $g_s$ (the string coupling constant) in the definition of the dilaton $\phi$ so that in the present paper the dilaton obeys $e^{\phi_{\mathrm{here}}}=g_se^{\phi_{\mathrm{there}}}$. Accordingly, the RR fluxes obey $g_sF_{p}^{\mathrm{here}}=F_{p}^{\mathrm{there}}$. 
Explicitly the dilaton is given by:
\eq{
e^{2\phi_{\mathrm{here}}}=\frac{g_s^2 e^{6A}}{1+\cos^2\!\theta ~\!e^{8A}}
~.
}
The forms $F_p$ have engineering dimensions $(\mathrm{length})^{p-1}$; the NSNS form $H$ and $\beta$ in eq.~(\ref{h}) have engineering dimensions $(\mathrm{length})^{2}$.
}
\eq{
\spl{\label{fluxes}
g_s F_0&=-L^{-1}\\
g_s F_2&=\frac{Le^{2C-4A}}{\cos\theta}~\!(1-\sin^2\!\theta ~\! e^{2(A-C)})~\!J+\dots\\
g_s F_4&=\frac{L^3e^{4C}}{2}~\!(1-2\sin^2\!\theta ~\! e^{2(A-C)})~\!J\wedge J+\dots\\
g_s F_6&=-\frac{3L^5e^{4C-2A}}{2}~\!\sin\theta~\! f(\theta)~\!J\wedge J\wedge(\d\psi+\mathcal{A})\wedge\d\theta
~.}
}
The ellipses in the second line above denote terms proportional to $(\d\psi+\mathcal{A})\wedge\d\theta$ ~\! whereas the ellipses in the third line above denote terms proportional to $J\wedge(\d\psi+\mathcal{A})\wedge\d\theta$~\!; we will not need the explicit form of these terms in the following. Let us also remark that there is no pole 
at $\theta=\pi/2$ in the expression for $F_2$, thanks to (\ref{exp2}).

It was shown in \cite{lt3} that, by virtue of the system of differential equations (\ref{7}), the RR and fluxes above obey the Bianchi identities
\eq{\label{bi}
\d F_p+ H\wedge F_{p-2}=0
~,}
where the NSNS three-form flux $H$ is closed, $\d H=0$. In fact for $F_0\neq 0$, as is the case for these massive vacua, $H$ is exact.  
Indeed, from the Bianchi identity for $F_2$, we see that $H$ can be written as:
\eq{\label{h}
H=\d(\beta-\frac{1}{F_0}F_2)~,
}
for some closed form $\beta$. Due to the gauge invariance of $H$, $\beta$ is  
only defined up to an exact form. Hence without loss of generality we may take 
$\beta$ to be a priori (i.e. before imposing the quantization of flux) an arbitrary real two-form in the second de Rham cohomology of the internal space, 
$\beta\in H^2(\mcal_6,\mathbb{R})$.

The RR forms in (\ref{fluxes}) are not closed, hence flux quantization cannot be directly imposed on them. However, the modified forms
\eq{\label{pf}
\tilde{F}_p:=(e^{\beta-\frac{1}{F_0}F_2}\wedge F)\vert_p
~,}
are closed by virtue of (\ref{h}) and the Bianchi identities (\ref{bi}). In the equation above $F$ is a RR polyform, i.e. a formal sum of RR forms of all orders: $F=\sum_pF_p$; on the right-hand side of (\ref{pf}) only the $p$-form contribution is selected. By flux quantization we mean imposing the conditions:
\eq{\label{pc}
\frac{1}{(2\pi\sqrt{\alpha'})^{p-1}}\int_{\mathcal{C}_i^{(p)}}\tilde{F}_p~\!\in\mathbb{Z}~,
}
where $\{\mathcal{C}_i^{(p)}\in H_p(\mcal_6,\mathbb{Z}), ~i=1,\dots, b_p(\mcal_6)\}$ is a set of $p$ cycles of the internal manifold furnishing a basis of integral $p$-homology.

For the following it will also be convenient to introduce the dimensionless quantity
\eq{
l:=\frac{L}{2\pi\sqrt{\alpha'}}~,
}
which is the AdS$_4$ radius in units of the string length.

$\bullet$ $p=0$

For $p=0$ the quantization condition (\ref{pc}) gives
\eq{\label{q1}
\frac{1}{g_sl}=n^{(0)}\in\mathbb{Z}
~,}
where we have taken (\ref{fluxes}) into account.

$\bullet$ $p=2$

As the basis of $H_2(\mcal_6,\mathbb{Z})$ we can take \cite{ms2} the set $\{\mathcal{C}_i^{(2)}\in H_2(\mathcal{B}_4,\mathbb{Z}), ~i=2,\dots, b_2(\mathcal{B}_4)+1\}$ (which constitutes a basis of integral second homology of the four-dimensional 
K\"{a}hler-Einstein base) at the point $\theta=0$ (the `north pole'), together with an $S^2$ cycle (the fiber):
\eq{\label{2cycle}
\mathcal{C}_1^{(2)}:=\{
(\psi,\theta):~0\leq\psi<\pi~,~~0\leq\theta\leq\pi
\}
~,}
at some fixed point of the base $\mathcal{B}_4$. Note that the second  Betti numbers of 
$\mathcal{B}_4$ and $\mcal_6$ are accordingly related through  $b_2(\mathcal{M}_6)=b_2(\mathcal{B}_4)+1$.

From (\ref{pf}) we find
\eq{\label{f2tilde}
\tilde{F}_2=F_0\beta
~.}
Then the quantization condition (\ref{pc}) gives  
\eq{\label{q2}
n^{(0)}d_i=n_i^{(2)}\in\mathbb{Z}
~,}
where we have defined
\eq{\label{bdef}
d_i:=\frac{1}{(2\pi\sqrt{\alpha'})^{2}}\int_{\mathcal{C}_i^{(2)}}\beta
~, ~~~i=1,\dots,b_2(\mathcal{M}_6)~.
}
It will be useful in the following to expand $\beta$ on a basis of 
the second de Rham cohomology of $\mcal_6$. Let $\{\omega^i\in H^2(\mcal_6,\mathbb{R})~,i=1,\dots,b_2(\mathcal{M}_6)\}$ be the basis dual to the cycles $\mathcal{C}_i^{(2)}$:
\eq{\label{dual}
\delta_i^j=\int_{\mathcal{C}_i^{(2)}}\omega^j
~, ~~~i,j=1,\dots,b_2(\mathcal{M}_6)
~.}
From the above and (\ref{bdef}) it then follows that
\eq{\label{bexp}
\frac{1}{(2\pi\sqrt{\alpha'})^2}~\!\beta=\sum_{i=1}^{b_2(\mcal_6)}d_i\omega^i
~.}

$\bullet$ $p=4$

As the basis of $H_4(\mcal_6,\mathbb{Z})$ we can take the set 
$\{\mathcal{C}_i^{(4)}:=\mathcal{C}_1^{(2)}\times\mathcal{C}_i^{(2)}~, ~i=2,\dots, b_2(\mathcal{M}_6)\}$ together with  $\mathcal{C}_1^{(4)}:=\mathcal{B}_4$ at the point $\theta=0$. The fourth  Betti number of $\mcal_6$ obeys: $b_4(\mathcal{M}_6)=b_2(\mathcal{M}_6)$, as it should by Poincaré duality.

From (\ref{pf}) we find
\eq{\label{4ta}
\tilde{F}_4=F_4-\frac{1}{2F_0}~\!F_2\wedge F_2+\frac{F_0}{2}~\!\beta\wedge\beta
~,}
while from (\ref{fluxes}) we calculate:
\eq{\label{4t}
F_4-\frac{1}{2F_0}~\!F_2\wedge F_2=-(2\pi\sqrt{\alpha'})^3~\!\d\big(
c(\theta) (d\psi+\mathcal{A})\wedge J
\big)~,
}
where we have defined
\eq{\label{ct}
c(\theta):=\frac{l^3e^{4C}}{2g_s}\left[
1-2\sin^2\theta~\! e^{2(A-C)}
+\frac{e^{-8A}}{\cos^2\theta}(1-\sin^2\theta~\! e^{2(A-C)})^2
\right]
~,
}
and we have taken (\ref{6}) into account\footnote{
From (\ref{fluxes}) we deduce that the left hand side of (\ref{4t}) is equal to $c(\theta)J\wedge J$ up to terms proportional to $\d\theta\wedge(d\psi+\mathcal{A})\wedge J$ (these are the terms which come from 
the ellipses in (\ref{fluxes})). On the other hand, it follows from the Bianchi identities that the left hand side of (\ref{4t}) is closed; taking that into account leads to eq.~(\ref{4t}).
}. 
Again, we note that  the expression above is regular 
at $\theta=\pi/2$ thanks to (\ref{attr}). 

We are now ready to impose the quantization condition (\ref{pc}). Integrating 
 (\ref{4ta}) over a four cycle $\mathcal{C}_i^{(4)}=\mathcal{C}_1^{(2)}\times\mathcal{C}_{i}^{(2)}$, with $i\neq 1$, we obtain:
\eq{\label{q3a}
\frac{~~\!l^3}{g_s}~\!v_if_1(A_0,C_0)-\frac{n_1^{(2)}n_i^{(2)}}{n^{(0)}}=n_{i}^{(4)}\in\mathbb{Z}
~,~~~i=2,\dots, b_2(\mathcal{M}_6)~,
}
where we have  taken (\ref{q1}),(\ref{q2}) into account
and we have defined
\eq{
\frac{~~\!l^3}{g_s}~\!f_1(A_0,C_0):=-\int_{\mathcal{C}_{1}^{(2)}} \d 
c(\theta) \wedge\d\psi=\pi[c(0)-c(\pi)]~,
}
so that $f_1$ is a function of the initial values $A_0$, $C_0$~\!; 
$v_i$ is the volume of the cycle $\mathcal{C}_{i}^{(2)}$, 
\eq{
v_i:=\int_{\mathcal{C}_{i}^{(2)}} J
~.}
Similarly, integrating (\ref{4ta})  
over the four-cycle $\mathcal{C}_1^{(4)}=\mathcal{B}_4$, taking (\ref{q1}),(\ref{q2}),(\ref{bexp}) into account, we obtain:
\eq{\label{q3b}
\frac{2l^3}{g_s}~\!\mathrm{vol}(\mathcal{B}_4)~\!e^{4(C_0-A_0)}\cosh(4A_0)
-\frac{1}{2n^{(0)}}\sum_{i,j\neq 1}{d^{ij}n_i^{(2)}n_j^{(2)}}=n_{1}^{(4)}\in\mathbb{Z}
~,
}
where we have defined the intersection numbers:
\eq{
d^{ij}:=\int_{\mathcal{B}_4}\omega^i\wedge\omega^j
~,}
and we have used the  fact that
\eq{
c(0)=\frac{~~\!l^3}{g_s}~\!e^{4(C_0-A_0)}\cosh(4A_0)~,
}
as follows from (\ref{ct}). 

Since $\tilde{F}_4$ is closed it can be expanded, up to an exact piece $\d\gamma_3$, on a basis of $H^4(\mcal_6,\mathbb{R})$. The latter may be taken to be generated by the forms $\{J\wedge J,~\omega_1\wedge\omega_i,~~\!i=2,\dots,b_2(\mathcal{M}_6)\}$. Taking (\ref{pc}) and Stokes' theorem into account, we  obtain the expansion:
\eq{\label{fexp}
\frac{1}{(2\pi\sqrt{\alpha'})^3}~\!\tilde{F}_4=\d\gamma_3+
\frac{n^{(4)}_1}{2\mathrm{vol}(\mathcal{B}_4)}~\!J\wedge J+
\omega_1\wedge \sum_{i=2}^{b_2(\mcal_6)}n^{(4)}_i\omega^i~.
}

$\bullet$ $p=6$

From (\ref{pf}) we obtain:
\eq{\label{pc6}
\tilde{F}_6=F_6-\frac{1}{F_0}F_2\wedge F_4+\frac{1}{3F_0^2}F_2^3+\beta\wedge \tilde{F}_4-\frac13 F_0~\!\beta^3~.
}
Integrating over $\mcal_6$,  taking (\ref{q1}),(\ref{q2}),(\ref{bexp}),(\ref{fexp}) into account, the quantization condition (\ref{pc}) gives
\eq{\label{q4}
\frac{l^5}{g_s}f_2(A_0,C_0)
+\frac{1}{n^{(0)}}\big( 
n_1^{(2)}n_1^{(4)}+\sum_{i,j\neq 1}{d^{ij}n_i^{(2)}n_j^{(4)}}
\big)
+\frac{n_1^{(2)}}{(n^{(0)})^2} 
\sum_{i,j\neq 1}{d^{ij}n_i^{(2)}n_j^{(2)}}
=n^{(6)}\in\mathbb{Z}~,
}
where $f_2(A_0,C_0)$ is a function of the initial conditions whose explicit form will not be necessary for the following.


\subsection{Solution of the quantization conditions}

The quantization conditions (\ref{q1}),(\ref{q2}),(\ref{q3a}),(\ref{q3b}),(\ref{q4}) can be solved for the 
parameters of the solution $g_s$, $l$, $A_0$, $C_0$ and $d_i$ (the moduli of the $\beta$ field) as follows. First we note that the conditions in (\ref{q2}) can be solved for the $d_i$'s. A priori (i.e. before charge quantization) the $d_i$'s are arbitrary 
real numbers; charge quantization imposes that they should be rational.

Next, the conditions (\ref{q1}),(\ref{q4}) can be solved for $g_s$, $l$. Note that if the AdS$_4$ radius is taken to be large in string units, $l>>1$, the string coupling $g_s$ will necessarily  be small, $g_s<<1$, by virtue of (\ref{q1}) which implies: 
\eq{
g_s\leq\frac{1}{l}
~.}
On the other hand, the ten-dimensional metric (\ref{1}) is proportional to $L^2$, hence the condition $l>>1$ implies small ten-dimensional curvature. 
In other words, the theory is weakly coupled in a region where the curvature is small so that the supergravity approximation can be trusted. 
%
%
These remarks are in agreement with the general conclusions of \cite{at} where it is argued that IIA supergravity cannot be strongly coupled in a region where the curvature is small.

Let us next consider the conditions in (\ref{q3a}). Eliminating the function $f_1$ we obtain the constraints:
\eq{\label{const}
\frac{v_i}{v_j}=
\frac{n^{(0)}n^{(4)}_i+n^{(2)}_1n^{(2)}_i}{n^{(0)}n^{(4)}_j+n^{(2)}_1n^{(2)}_j}
~,~~~i,j\neq 1~,
}
which imply that the volumes $v_i$ must be rationally related. This is indeed the case \cite{ms2}, as can be seen from the fact that the $v_i$'s are (up to an overall normalization) Chern numbers of the anticanonical bundle of $\mathcal{B}_4$. Given the set of rational numbers ${v_i}/{v_j}$, which are fixed by the topology of $\mcal_6$, the constraints (\ref{const}) can easily be solved by (infinitely many different) suitable choices of the integers $n^{(2)}_i$, $n^{(4)}_i$. 

Assuming the constraints (\ref{const}) have been solved,  (\ref{q3a}) reduces to a single independent equation for $f_1$:
\eq{\label{q3aa}
v_2f_1(A_0,C_0)-\frac{n_1^{(2)}n_2^{(2)}}{n^{(0)}}=n_{2}^{(4)}
~.}
Hence the quantization conditions (\ref{q3a}),(\ref{q3b}) amount to two  equations for the two unknowns $A_0$, $C_0$. This completes the solution of the flux quantization conditions.

{}Furthermore we can show that the initial values $A_0$, $C_0$ can be tuned independently. For that it suffices to show that $n^{(4)}_1$, $n^{(4)}_2$, i.e. the right-hand sides of eqs.(\ref{q3b}),(\ref{q3aa}) (whose solution determines $A_0$, $C_0$), can take on any (integer) value. Therefore it suffices to show that the constraints (\ref{const}) can be solved for any  $n^{(4)}_2$.  This is indeed the case: Let $v_2/v_i=p_i/q_i$, for some integers $p_i$, $q_i$.  
Then it can easily be seen that one possible solution to (\ref{const}) is given by
\eq{\spl{
n^{(2)}_1=n^{(0)}~,~~~n^{(2)}_2&=\prod_{i=3}^{b_2(\mcal_6)}p_i- n^{(4)}_2~,~~~\\
n^{(4)}_i=\frac{q_i}{p_i}\prod_{j=3}^{b_2(\mcal_6)}p_j-n^{(2)}_i~&,~~~~~i=3,\dots,
b_2(\mcal_6)
~,}}
which solves for $n^{(2)}_1$, $n^{(2)}_2$, $n^{(4)}_i$, $i\geq3$, leaving   $n^{(4)}_2$ as an independent parameter.

In conclusion: the quantization conditions can be solved in a regime where 
$g_s$ and the curvature are arbitrarily small, so that the supergravity solution 
can be trusted. Moreover $A_0$, $C_0$ can take on independent (discrete) values.

\section{Kaluza-Klein spectrum}\label{kk}

The interest of vacua with scale separation is linked to 
the decoupling of the KK tower of massive modes. 
The latter problem could in principle be studied systematically 
for the vacua of \cite{lt3}, although this is beyond the scope 
of the present paper. Instead we will examine here the  
KK spectrum in the simplified 
case of a massive ten-dimensional scalar.

Consider a ten-dimensional 
massive  scalar $\Phi$ in the geometry given by (\ref{1}),(\ref{3}),\footnote{The analysis of this section is presented for the 
ten-dimensional metric in the string frame. It is straightforward to show that the same conclusions hold, and in particular the asymptotic formula (\ref{asymtext}) remains valid, in any frame.}
\eq{\label{10ds}
(-\Delta_{10}+M^2)\Phi(x,y)=0
~,}
where $\Delta_{10}$ is the ten-dimensional Laplacian, and $x$, $y$ are coordinates of AdS$_4$, $\mcal_6$ respectively. 
Let us further decompose:
\eq{
\Phi(x,y)=\sum_n\varphi_n(x)\omega_n(y)
~,}
where the $\omega_n(y)$'s are orthonormal weighted eigenfunctions 
with weight $e^{-2A}$ of a modified Laplacian of  $\mcal_6$ to eigenvalues $\lambda_n$:
\eq{\label{int}
-\hat{\Delta}_6\omega_n(y)=\lambda_ne^{-2A}\omega_n(y)
~,}
where the operator $\hat{\Delta}_6$ is defined through:
\eq{
\hat{\Delta}_6\omega:=\left({\Delta}_6 +4\partial A\cdot\partial 
-L^2M^2 \right)\omega
~,}
with ${\Delta}_6$ the Laplacian of  $\mcal_6$. 
It then follows that the four-dimensional Kaluza-Klein modes $\varphi_n(x)$ obey
\eq{\label{kg}
-\Delta_4\varphi_n(x)+ \frac{\lambda_n}{L^2}~\!\varphi_n(x)=0
~,}
where $\Delta_4$ is the Laplacian of AdS$_4$ (with radius $L$). 
For an internal space of typical size $L_{\mathrm{int}}$ we expect the KK masses $m_n^2$ to scale  as: 
\eq{\label{decoup}
m_n^2:= \frac{\lambda_n}{L^2}\sim  \frac{n^2}{L_{\mathrm{int}}^2} 
~.}
Let us now turn to the study of (\ref{int}) and the spectrum of eigenvalues $\lambda_n$. In the following 
it will suffice to look at the case where  $\omega_n(y)$ does not depend on the 
coordinate $\psi$, a choice which we make for simplicity. Let us  further decompose:
\eq{
\omega_n(y)=\chi(w)g_n(\theta)
~,}
where $w$ is a coordinate of $\mathcal{B}_4$ and $\chi(w)$ is an eigenmode of the Laplacian of $\mathcal{B}_4$,
\eq{
-\Delta_{\mathcal{B}_4}\chi(w)=\mu \chi(w)
~,}
 to eigenvalue $\mu$. 
Taking the form of (\ref{3}) into account,  (\ref{int}) 
reduces to the following second order ordinary differential 
equation for $g_n(\theta)$ 
(where a prime denotes 
differentiation with respect to $\theta$):
\eq{\label{sl1}
\big(pg_n'\big)'
+(\lambda_n q-r)g_n=0 ~,}
where:
\eq{\label{coeffs}
p(\theta):=\frac{e^{4(A+C)}}{f}~\!\sin\theta~,~~~
q(\theta):= e^{4(A+C)}f\sin\theta ~,~~~
r(\theta):= e^{6A+2C}f\sin\theta~\!(\mu+L^2M^2e^{2C}) ~,
}
and $f(\theta)$ was given in (\ref{4}). 

Eq.~(\ref{sl1}) is a singular Sturm-Liouville (SL) problem (see appendix \ref{sl} for more details), since $p$ vanishes linearly at the endpoints $0$, $\pi$ of the $\theta$-interval, as we can see from (\ref{exp1}). However, it can be mapped to a regular SL problem given by (\ref{sl2ap}),(\ref{trnsf}) by means of the transformation $h_n:=g_n/u$ 
where the function $u$ is given by:
\eq{\label{udef}
u=1-\log(\sin\theta)~.
}
Indeed the function $u$ defined above is strictly positive on the interval $I=(0,\pi)$ as required by the lemma in appendix \ref{sl}. Moreover, for $u$ as given above the coefficients $\tilde{p}$, $\tilde{q}$, $\tilde{r}$ defined in (\ref{trnsf}) obey: $1/\tilde{p}$, $\tilde{q}$, $\tilde{r}\in L(I,\mathbb{R})$, as required for a regular SL problem. To see that $1/\tilde{p}$ is integrable note that 
$1/\tilde{p}\sim 1/(\theta(\log\theta)^2)$ for $\theta\sim 0$ and similarly for $\theta\sim\pi$, as follows from the expansion (\ref{exp1}). By the same reasoning it easy to see that $\tilde{q}$, $\tilde{r}$ are also integrable.

The upshot of the preceding analysis is that we have mapped the spectrum of 
KK masses to the eigenvalues of a regular SL problem with $\tilde{p}>0$ on $I$. From (\ref{asym}) we then obtain the following asymptotic formula for the eigenvalues:
\eq{\label{asymtext}
\frac{\lambda_n}{n^2}\rightarrow c:=\pi^2\left(\int_0^{\pi}f\d\theta\right)^{-2}~,
~~ \mathrm{as}~~n\rightarrow\infty
~,}
where we took into account that $\tilde{q}/\tilde{p}=f^2$ as follows from (\ref{coeffs}),(\ref{trnsf}). Moreover it follows from  eq.~(\ref{bound}) that $c\leq 4$. In other words, ${\lambda_n}/{n^2}$ is at most of order one for large $n$ thus by comparing with 
(\ref{decoup}) we see that the asymptotic KK spectrum is governed by an effective KK scale of at least the order of the AdS$_4$ radius: 
$L\lesssim L_{\mathrm{int}}$.

\section{Conclusions}\label{conclusions}

The absence of examples of supersymmetric pure-flux AdS vacua with scale separation has led to a folk belief that there is a no-go which excludes them. 
For the reasons discussed in the introduction, if such a no-go does not exist  supersymmetric pure-flux AdS vacua with scale 
separation would provide highly desirable starting points for realistic 
compactifications of string theory.

Supersymmetric pure-flux AdS vacua have been around since the first days of supergravity. In these early Freund-Rubin type vacua  \cite{fr,dp,np} the absence of scale separation is straightforward: the radius of curvature of the internal space is everywhere of the same order as the radius of curvature of the AdS space. As we have seen in section \ref{sec3}, the same is true for the more recent vacua of the type of \cite{lt}. These include the explicit examples of \cite{bc,to}; the exhaustive list of explicit examples known to date can be found in \cite{kt}.

In contrast, the question of scale separation in the case of the AdS vacua of \cite{lt3} examined in the present paper is much more subtle. As we have seen although the geometric moduli of the solution can be tuned to ensure an arbitrarily large hierarchy of scales in some neighborhood of the space, there is an attractor mechanism at work which forces all scales to be of the same order in some other neighborhood. The scalar curvature of the internal space exhibits a similar behavior: although it can be tuned to be much larger than the external scalar curvature in some neighborhood of the space, the ratio of the two curvatures is necessarily of order one in some other neighborhood.

Furthermore, the analysis of the KK spectrum of a ten-dimensional scalar shows that the higher modes behave as if the effective size of the internal space were of the same order as the AdS$_4$ radius. We should stress however that this does not necessarily imply that there is no separation of scales: it remains a logical possibility that the low-lying KK modes `see' an effective internal space size which is much smaller than the AdS$_4$ radius. In this scenario the low-lying KK modes would scale as $m_n^2\sim n^2/L_{\mathrm{low}}^2$ while the higher modes scale as $m_n^2\sim n^2/L_{\mathrm{high}}^2$, with $L_{\mathrm{low}}<<L_{\mathrm{high}}\sim L$. 
In principle, the  matter could be settled by a systematic derivation of the KK spectrum. What makes  this direct approach complicated at present is the fact that the functions $A(\theta)$, $C(\theta)$ are only known implicitly as solutions of the system of first-order differential equations (\ref{7}).

\appendix

\section{$S^2(\mathcal{B}_4)$ bundles}\label{details}

In this appendix we summarize some relevant facts about the geometry of the 
sphere bundles $S^2(\mathcal{B}_4)$. 
The range of $\psi$ in (\ref{3}) follows from 
the requirement that the $S^2$ fiber should be smooth: indeed fixing a point on the $\mathcal{B}_4$ base, the  $\psi$ coordinate parameterizes a circle fibered over the interval $\theta\in [0,\pi]$. Moreover, as can be seen from (\ref{3}),(\ref{4}), the circle smoothly shrinks to zero at the endpoints of the interval provided $\psi$ has period $\pi$.

Furthermore let $\mathcal{P}$ be the connection 
 on the anticanonical bundle of $\mathcal{B}_4$. It follows that 
\eq{
\d\mathcal{P}=-\mathcal{R}~,
}
where $\mathcal{R}$ is the 
Ricci form of $\mathcal{B}_4$. On the other hand, for a K\"{a}hler manifold the 
Ricci form $\mathcal{R}$ is related to the Ricci tensor via:
\eq{
{R}_{mn}=-\mathcal{I}_m{}^k\mathcal{R}_{kn}~,
}
where $J$ is the K\"{a}hler form and $\mathcal{I}_m{}^k:=J_{mn}g^{nk}$ is the 
complex structure. Moreover, for a  K\"{a}hler-Einstein manifold 
we have 
\eq{
{R}_{mn}=\Lambda {g}_{mn}~,
}
where the `cosmological constant' $\Lambda$ is given by $\frac{1}{d}R$ for a 
$d$-dimensional space. Combining the above we obtain
\eq{
\d\mathcal{P}=-\Lambda J~.
}
In the case of $\mathcal{B}_4$ we have $\Lambda=2$ from eq.~(\ref{5}) and therefore by comparing the equation above to (\ref{6}) we obtain:
\eq{
\mathcal{A}=\frac{1}{2}\mathcal{P}~.
}
Hence if we set $(\d\psi+\mathcal{A})=\frac{1}{2}(\d\tilde{\psi}+\mathcal{P})$, then $\tilde{\psi}$ has twice the period $\psi$. Moreover, since $\psi$ is the circle coordinate in the anticanonical bundle of $\mathcal{B}_4$, $\tilde{\psi}$ must have period $2\pi$ and so the period of $\psi$ is $\pi$. As seen previously, this  consistent with the range one gets by requiring that the $S^2$ fiber be smooth.

\section{Sturm-Liouville}\label{sl}

Sturm-Liouville theory is a well-established subject about which many standard textbooks are available. Here we review some results which are used in the main text. Our exposition follows closely \cite{zbook}.

We will denote an open interval by $(a,b)$, with $-\infty\leq a<b\leq\infty$; $[a,b]$ denotes the closed integral which includes the endpoints $a$, $b$, regardless of whether these are finite or infinite. 
Let $I$ be any interval of the real line, open, closed, bounded or unbounded. 
By $L(I,\mathbb{R})$ we will denote the space of real functions $g(\theta)$ defined for $\theta\in I$, such that
\eq{
\int_a^b|g(\theta)|\d \theta<\infty~.
}
We define a regular SL problem as consisting  of the 
differential equation (where a prime denotes 
differentiation with respect to $\theta$):
\eq{\label{sl1ap}
\big(pg'\big)'
+(\lambda q-r)g=0
~~ \mathrm{on}~~I=(a,b), ~~-\infty\leq a<b\leq\infty ~,}
with real coefficients $p$, $q$, $r$ satisfying
\eq{\label{slcoeffs}
p>0~~ \mathrm{on}~~I,~~ \mathrm{and}~~ \frac{1}{p},q,r\in L(I,\mathbb{R})~~;
}
and either `separated' boundary conditions:
\eq{\spl{\label{sep}
A_1&g(a)+A_2(pg')(a)=0, ~~ A_1^2+A_2^2\neq 0\\
B_1&g(b)~\!+B_2(pg')(b)=0, ~~ B_1^2+B_2^2\neq 0
~,}}
for some real numbers $A_1$, $A_2$, $B_1$, $B_2$, or `periodic' boundary conditions:
\eq{\label{a5}
g(a)=g(b)~,~~~(pg')(a)=(pg')(b)
~.}
Some of the conditions in the definition above can be significantly weakened, but this will not be necessary for our purposes.

It is well-known that the regular SL problem defined in (\ref{sl1ap})-(\ref{a5}) above has solutions $g_n(\theta)$ only for certain values of $\lambda=\lambda_n$. Specifically, 
we have the following theorem (see e.g. \cite{zbook}, p.72):
\begin{enumerate}
\item The eigenvalues are bounded below and can be ordered to 
satisfy:
\eq{\label{eigen}
-\infty<\lambda_0\leq\lambda_1\leq\lambda_2\leq\dots;~~\lambda_n\rightarrow
+\infty~,~~ \mathrm{as}~~n\rightarrow\infty
~.}
Each eigenvalue may be simple or double (each eigenvalue is simple if the boundary conditions are separated) but there cannot be two consecutive equalities in (\ref{eigen}) since for any value of $\lambda$ equation (\ref{sl1ap}) has exactly two linearly independent solutions. 
\item The following asymptotic formula holds:
\eq{\label{asym}
\frac{\lambda_n}{n^2}\rightarrow \pi^2\left(\int_a^b\sqrt{\frac{q}{p}}\right)^{-2}~,
~~ \mathrm{as}~~n\rightarrow\infty
~.}
\end{enumerate}

The following Lemma \cite{zpaper} is used in section \ref{kk}. Suppose $u$ is real function  such that $u>0$ on $I$, then $g$ is a solution of (\ref{sl1ap}) to eigenvalue $\lambda$ if and only if $h:=g/u$ is a solution of the following SL problem to the same eigenvalue $\lambda$:
\eq{\label{sl2ap}
\big(\tilde{p}h'\big)'
+(\lambda \tilde{q}-\tilde{r})h=0
~,}
where
\eq{\label{trnsf}
\tilde{p}:=pu^2~,~~~ \tilde{q}:=qu^2~,~~~ \tilde{r}:=ru^2-u\big(pu'\big)'~.}
The above follows by direct calculation.

\vfill\break

\end{document}